\documentstyle[multicol,prl,aps]{revtex}
\input{epsf.tex}
\date{\today}
\begin{document}

\title{Strong Coupling and Double Gap Density of States in Superconducting
MgB$_{2}$}

\author{F. Giubileo$^\dagger$, D. Roditchev, W. Sacks, R. Lamy and J. Klein}

\address{Groupe de Physique des Solides, Universit\'es Paris 7 et Paris
6, \\ Unit\'e Mixte de Recherche C.N.R.S. (UMR 75\ 88), 2 Place
Jussieu, 75251 Paris Cedex 5, France}

\address{$^\dagger$Physics Department and INFM Unit, University of Salerno,
via S. Allende, 84081 Baronissi (SA), Italy}

\maketitle

\begin{abstract}
Using scanning tunneling spectroscopy at $T = 4.2\,$K, we perform
simultaneously the topographic imaging and the quasiparticle
density of states (DOS) mapping in granular MgB$_{2}$. We observe
a new type of spectrum, showing a pronounced double gap, with the
magnitudes of $\Delta_{S} = 3.9$ meV and $\Delta_{L} = 7.5\,$ meV,
i.e. well below and well above the BCS limit. The largest gap
value gives the ratio $2\Delta_{L}/k_{B} T_{c} = 4.5$, which
implies strong electron-phonon coupling. Other superconducting
regions are found to have a characteristic BCS-shaped DOS.
However, the variation of the spectral shape and lower gap widths,
from 2.0 meV to 6.5 meV, indicate the importance of surface
inhomogeneity and proximity effects in previously published
tunneling data. Our finding gives no evidence for any important
gap anisotropy. Instead, it strongly supports the multiple gap
scenario in MgB$_{2}$ in the clean limit, and the single gap
scenario in the dirty limit.
\end{abstract}

\vskip 2mm

{\small \ \ \ \ \ PACS numbers: 74.20.-z, 74.50.+r, 74.70.Ad}

\vskip 3mm


\begin{multicols}{2}

Since its announcement by Nagamatsu {\it et al.} in January 2001 \cite{Nagamatsu}, the
discovery of superconductivity in the simple metallic compound
MgB$_{2}$ has stimulated a great effort in the field. A very high
critical temperature $T_{C} \simeq 39\,$K, well above the values reported
for all other conventional superconductors, and quite simple binary
structure, make this material a good candidate for technological
applications.  At the same time this unusually high $T_{C}$ opens a new
fascinating question about the electronic structure and pairing mechanism
involved.

Numerous experimental reports confirm that MgB$_{2}$ is a BCS-type
phonon-mediated superconductor.  A significant boron isotope
effect ($\, \alpha \simeq 0.26\,$) was observed in both
magnetization and specific heat measurements \cite{Budko}.
Measurements on nuclear spin-lattice relaxion rate in the $
^{11}$B NMR study by Kotegawa et al.  \cite{Kotegawa} have shown
the existence of an s-wave superconducting state with a large
isotropic gap ($2 \Delta /k_{B}T_{C} \simeq 5$) well above the
weak-coupling BCS value of 3.52.  The strong coupling scenario
could be explained considering the high phonon frequencies
($\omega \simeq$ 700 K) in MgB$_{2}$ due to the light boron mass
\cite{Kortus}. To complicate the picture, it has been shown
theoretically that MgB$_{2}$ has a particular band structure in
which both two- and three-dimensional bands are present and that,
in the clean limit, could even give rise to a multiple gap
superconductivity \cite{Liu}.

The direct measurement of the superconducting DOS of MgB$_{2}$ is
urgently needed. Photoemission experiments have already been performed
\cite{Takahashi,Tsuda} but are, unfortunately, limited by energy
resolution. Details of the magnitude and
symmetry of the superconducting gap can be obtained with
tunneling spectroscopy (TS), giving directly the quasiparticle DOS with an
energy resolution of a few kT.  If the full scanning TS (STS) is
performed, the spatial mapping of the local DOS is possible. It has
also proved useful in the case of inhomogeneous systems \cite{Cren}.
Finally, in the tunneling conductance, spectral features found above the gap
allow a self-consistent analysis of the electron-phonon
interaction \cite{RowellMac}.

Unfortunately, the published tunneling spectra clearly diverge
\cite{Rubio,Karapetrov,Schmidt,Sharoni,Zaza} and no STS data have yet been
presented. While the shapes of the spectra indicate an isotropic gap for the pairing
symmetry, the gap widths are quite different, all being too small
to be realistic for the bulk MgB$_{2}$.  Moreover, in
\cite{Karapetrov} the spectra were found exceedingly smeared.  To
explain these different results, even a very sophisticated symmetry was
suggested for an anisotropic order parameter \cite{Chen}.
In this series tunneling experiments
\cite{Rubio,Karapetrov,Schmidt,Sharoni,Zaza}
no evidence whatever for a multiple (or double) gap scenario was shown.
Consequently, an important uncertainty exists in establishing
the shape of the quasiparticle DOS and the gap magnitude
pertinent to the bulk superconductivity.

In this Letter we show that the complete STS, even on the small scale
of a few hundred nanometers, allows one to interpret the wide
variation in the previous tunneling data. The measurements were performed
on MgB$_{2}$ powder ($T_{C}=38.7\,$K), which was glued to the
sample holder by silver paint and mechanically etched Pt/Ir wires
were used as STM tips.  Measured locally, the conductance spectra reveal
a distinct double gap structure as shown in Fig.1.
These show little or no smearing other than thermal broadening at 4.2 K.
As we shall argue, this double gap is an intrinsic property of MgB$_{2}$
and not a proximity effect \cite{Mcmillan}.

The tunneling conductance in Fig.1 is also characterized by a very
flat spectral background, good sharpness, conservation of
quasiparticle states, and by no shape dependence on the tip-sample
distance. These conditions demonstrate the good quality of the
tunneling junction. The spectra show a relatively small gap
$\simeq 5$ mV with almost no excitations inside.  It is followed
by two further peaks, that are seen in both occupied and empty
states sides. As we shall describe below, these features
correspond to the larger second superconducting gap $\Delta = 7.5
$ meV. It exceeds all those reported in the tunneling literature
\cite{Rubio,Karapetrov,Schmidt,Sharoni,Zaza}. Consequently, the
corresponding ratio $2\Delta /k_{B}T_{C} = 4.5$ clearly points to
strong electron-phonon coupling.

\begin{figure}[h]
\vbox to 4.5cm{
   \epsfxsize=7.5 cm
   \epsfbox{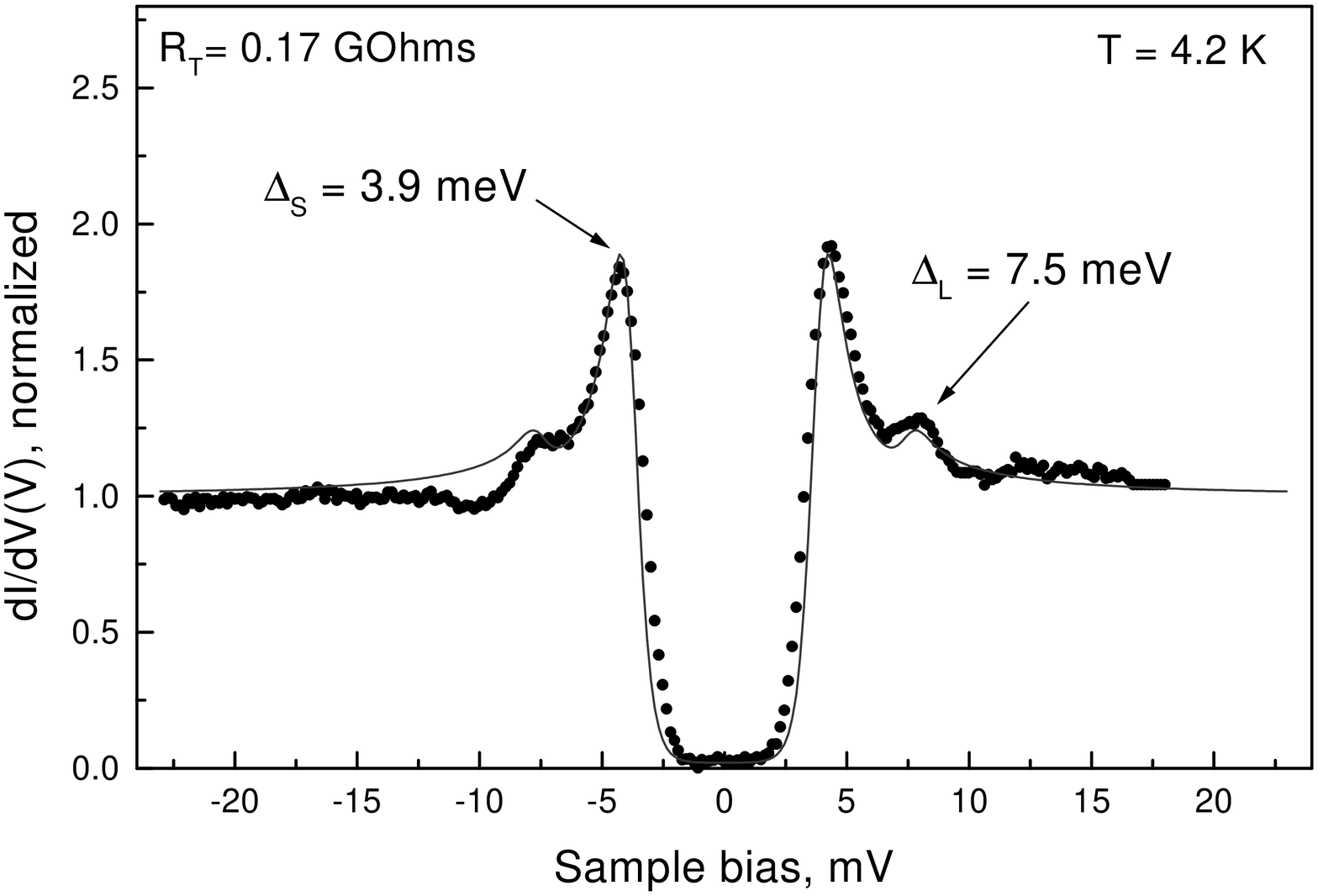}
}
  \label{tre}
\end{figure}

{\small FIG.1\ : Double gap structure in local conductance spectra (dots).  A
   small gap $\Delta_{S} = 3.9$ meV and a large one $\Delta_{L} = 7.5$
   meV exist together.  Flat spectral background and zero states at the Fermi energy are
   characteristic features of such spectra.
   Solid line: A generalized BCS fit, using Eqs. (1-3), with the parameters $\Gamma = 0$ meV
   and spectral weights $C_{S} = 0.94$ and $C_{L} = 1- C_{S} = 0.06$.}

\vskip 1 mm

An analogous double-gap spectrum was very recently observed in an
inverted SIN junction \cite{giubileo} and in an
independent point-contact spectroscopy experiment \cite{Szabo}. It has
also been shown that both gaps close at the same temperature
\cite{giubileo}, i.e. the critical temperature $T_{C}$ of the bulk material. Thus both
gaps are intimately related to the superconducting properties of
MgB$_{2}$. This behavior is consistent with the theoretical prediction
of Liu et al. \cite{Liu} for two gap superconductivity in the clean limit for
bulk MgB$_{2}$.

A double gap, such as in Fig.1, is remarkable
but could {\it a priori} originate from different
physical scenarios. In the first one, the small gap reflects a
proximity superconductivity induced in a thin metallic surface
layer.  In this case, the observed bump indicates the
position of the superconducting gap of the bulk MgB$_{2}$
\cite{Mcmillan}.  In the second scenario the small gap is due to a
weakened superconductivity on the very surface. Then two
tunneling contributions may exist: one corresponding to tunneling to the
weakly superconducting surface  and another to the bulk MgB$_{2}$
directly. Again, the feature at 7.5 mV determines the
energy of the bulk gap.

A third and simpler interpretation is the intrinsic double gap scenario
for the bulk superconductivity in MgB$_{2}$, as predicted for the clean limit
\cite{Liu}. It originates from the existence of two
different sheets of the Fermi surface, quasi-2D cylinders and a
three-dimensional tubular network, and from a very particular
anisotropic phonon spectrum.  This results in a strong
electron-phonon coupling with the quasi-2D bands and a much weaker
coupling to the 3D band.  Finally, two distinct and observable
superconducting gaps appear in the quasiparticle spectrum,
one larger and the other smaller than the BCS value.
The effect is very sensitive to impurities,
and vanishes in the dirty limit in which just one gap exists.
Taking into account the very short coherence length $\xi \leq 50\,$ \AA\cite{Finnemore},
it is quite possible that both clean and
dirty limit conditions may be found on the surface of such a granular
sample.

Further tunneling data are presented in Fig.2 as strong evidence in
favor of two gap superconductivity. Here the
conductance spectra, selected from different locations of the
same sample, clearly show double-peak structures of various relative
amplitudes. Remarkably, the energy positions of the gaps remain fixed,
which is difficult to explain in terms of a proximity effect. Indeed, in the
latter case the DOS is a complicated function of the SN sandwich
parameters (interface transparency, metallic layer thickness,
diffusion coefficient, etc.) and may be calculated using Usadel's
equations \cite{Usadel}. In particular, both the apparent amplitudes of the
peaks and their energy positions (especially those of the small
induced gap) are very sensitive to any change of the SN sandwich
parameters. This is evidently not the case in Fig.2 (a), in which the
relative amplitude of the gaps changes markedly, but their
positions remain robust, clearly revealing two characteristic
energies.

\vskip 0.2 cm

\begin{figure}[]
    \epsfxsize=8 cm
    \epsfbox{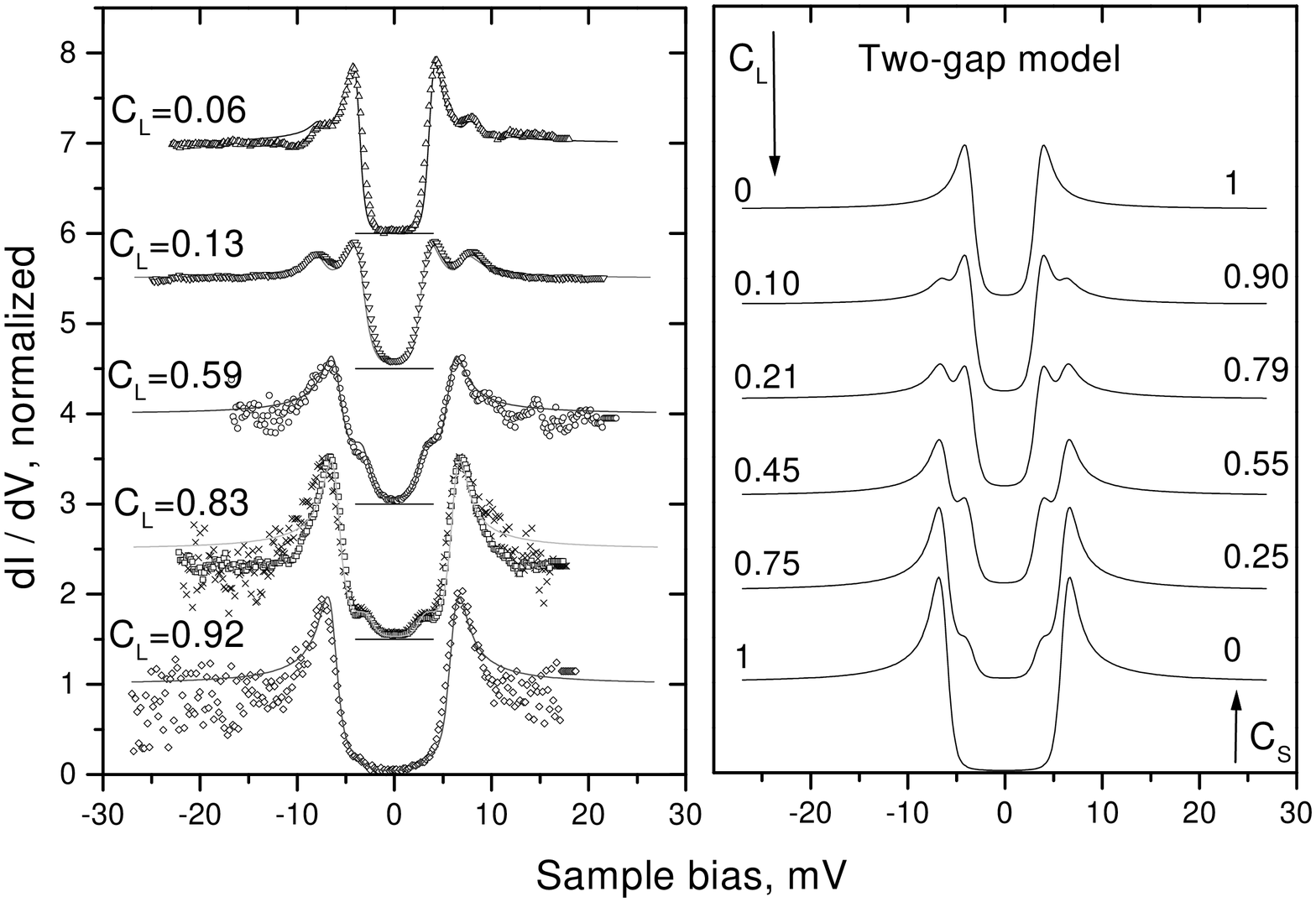}

 \label{four}
\end{figure}

{\small FIG.2. (a) Tunneling conductance spectra (T = 4.2 K) normalized to unity and shifted for
 clarity (horizontal bars indicate zero conductance). The values of
 the gap widths ($\Delta_{S}$ and $\Delta_{L}$) are fixed but
 the relative strength of the peaks vary.
 The spectra can be fitted by only adjusting the coefficient $C_{L}$
 the weight-factor of the large gap to the total density of states.
 The spectrum $C_{L}=0.13$ was obtained in inverted SIN case
 \cite{giubileo}.
 (b) Theoretical curves for the two-gap model, Eqs. (1-3), with
 different values for the tunneling probabilities $C_{L}$,
 $C_{S}$=1-$C_{L}$.
 The case $C_{L}$=1 gives the usual BCS density of states
 with gap width $\Delta_{L}$, while $C_{L}$=0 ($C_{S}$=1)
 represents the BCS case with gap width $\Delta_{S}$.}

\vskip 1mm
The double-gap superconductivity
explains naturally the above observations. While the peak positions are
fixed, i.e. corresponding to two energy gaps,
the spectral weight of each reflects the
probability of tunneling to different electronic bands (2D and
3D). Indeed, in powder based samples the orientation of grains with respect to
the tunneling cone varies, as do the respective tunneling
probabilities, from one location to another. This explains the
variation of the relative amplitudes of the coherence peaks from
grain to grain. As illustrated in Fig. 2 (b), the double-gap spectra can be generated
by a weighted sum of two BCS density of states with
coefficients $C_{L}$ and $C_{S}$ ($C_{L} + C_{S} = 1$) for the large
and small gap DOS, respectively\ :
\begin{equation}
\sigma(V) = C_{L}\ \sigma^{L}(V) + C_{S}\ \sigma^{S}(V)\ ,
\label{cond}
\end{equation}
with \ :

\begin{equation}
\sigma^{L,S}(V)\propto\int_{-\infty }^{+\infty }g(\epsilon-eV)\
\rho^{L,S}(\epsilon)\ d\epsilon \ ,
\end{equation}

\begin{equation}
\rho^{L,S}(\epsilon)={\rm Re}\left[\frac{\epsilon- {\rm
i}\Gamma}{\sqrt{(\epsilon-{\rm i} \Gamma)^{2}-\Delta_{L,S}^{2}} }
\right] \ .
\end{equation}
Here $g(\epsilon)=-\partial f(\epsilon)/\partial \epsilon$ is the usual
thermal broadening function and $\Gamma$ is Dyne's smoothing parameter.

This simple analysis, Eqs. (1-3), appears sufficient to fit
all of the double-gap conductance spectra of Fig. 2 (a) (solid lines in the figure).
We obtain statistically, $\Delta_{L}= 7.5 \pm 0.5$ meV and
$\Delta_{S} = 3.5 \pm 0.4$ meV with the smearing term not exceeding
$\Gamma = 0.2$ meV.  The double structure and other DOS features
(zero bias conductance, spectral background, etc.) are perfectly reproduced
by the theoretical function for different values of $C_{L}$.
Even for the bottom spectrum in
Fig.2 (a), for which the existence of the inner gap is not so evident,
it was necessary to include both gap contributions.

The double gap structure represents only a very small fraction of
the acquired data over the surface. The STS technique allows to
systematically explore different regions showing a variety of
quasiparticle spectra. A complete $I(V)$ spectrum is acquired
locally at each pixel $(x,y)$ of the STM image, so that a perfect
correspondence between topographic and spectroscopic information
is achieved. Moreover, a huge number of tunneling spectra are
acquired (up to 65536 $I(V)$ curves per image) allowing a
significant statistics. The raw $I(V)$ curves are numerically
differentiated and presented either as a series of tunneling
conductance maps $dI/dV(V_{i},x,y)$, at a chosen bias $V_{i}$, or
as tunneling conductance spectra $dI/dV(V,x_{i},y_{i})$ measured
in particular locations $(x_{i},y_{i})$. More details concerning
our STS experimental setup can be found in\cite{Cren}.

In Fig.3  we give an example of such an STS measurement. Image (a)
is the topography, acquired simultaneously, of the surface (220 $\times$ 220
nm$^{2}$) showing relatively flat
terraces, with a roughness of about 1-3 nm
separated by steps 1-20 nm high. It does happen that the tip jumps from one
region to another due to the surface roughness on a larger scale.
A single occurence of this is indicated in (a) by the arrow.

\begin{figure}[h]
\centering \vbox to 5.5 cm{ \epsfxsize=6.0 cm \epsfbox{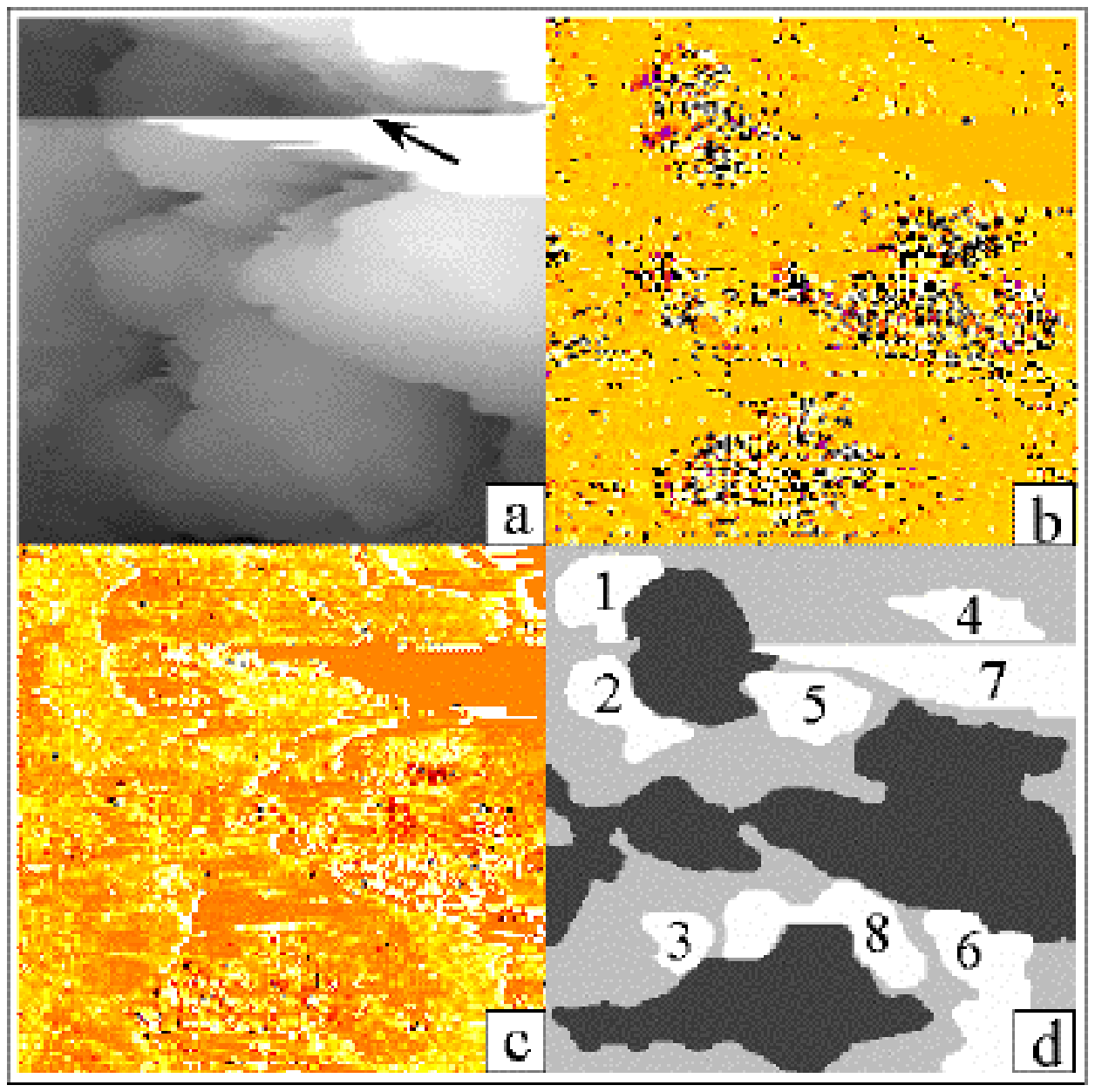}}
\label{uno}
\end{figure}

{\small FIG. 3. Complete DOS mapping of a $220 {\rm nm} \times 220 {\rm nm}$
area of a chosen MgB$_{2}$ grain. (a) Topographic image acquired
in the constant current mode (sample bias V = -18 mV, I = 80 pA).
Flat regions are resolved.  Black arrow indicates a tip jump
during a scan.  (b) Conductance map of the same area at V = -15
mV.  (c) Conductance map at V = +5 mV.  Bright and dark zones
respectively indicate high and low conductance zones.  (d) In
black: `contaminated' regions (as found from (b) ); in white:
selected regions (numbered from 1 to 8) described in the text.}

\vskip .5mm

In Fig.3 (b) we show the tunneling conductance map at $V = -15$ mV
proving that the DOS is spatially inhomogeneous. Regions in which
high-quality tunneling conditions are not met (the current is
noisier, probably due to the surface contamination) are in black
in Fig.3 (d). Of particular interest is the conductance map for $V
= +5$ mV, i.e. at an energy close to the expected superconducting
gap edge, Fig.3 (c). In this map, bright and dark zones reveal the
co-existing regions with high and low conductances.  Comparing the
topographic image (a) with the two conductance maps (b) and (c),
we can easily select the most `significant' regions in which
noiseless tunneling conditions are achieved on a relatively flat
surface.  Such regions, of a typical size of 10-30 nm, are colored
in white and numbered from 1 to 8 in Fig.3 (d). As a simple
approach, the spectra are averaged over each selected region and
the results are plotted in Fig.4.

The conductance spectra from the regions 1 and 2 are quite similar
and both spectra present a smoothed reduced gap of about 2.5 meV.
The ratio $2\Delta /k_{B}T_{C} = 1.48$ ($T_{C}$ = 39\,K for the
bulk) is less than two times smaller than the weak coupling BCS
ratio.  Such spectra originate either from the weakened $T_{C}$ on
the surface (due to the local surface degradation) or from a
proximity induced gap.  The conductance at zero-bias is far from
zero and these spectra show a strong smearing term $\Gamma =
0.3\,\Delta$.  Contrary to what was suggested in
\cite{Karapetrov}, $\Gamma$ does not reflect the properties of the
bulk MgB$_{2}$ since it varies spatially, and even vanishes in
other regions. This type of spectrum was also observed by Sharoni
et al. \cite{Sharoni} and explained in terms of the proximity
effect.

\vskip 0.2 cm

\begin{figure}[h]
\epsfxsize=7.0 cm \epsfbox{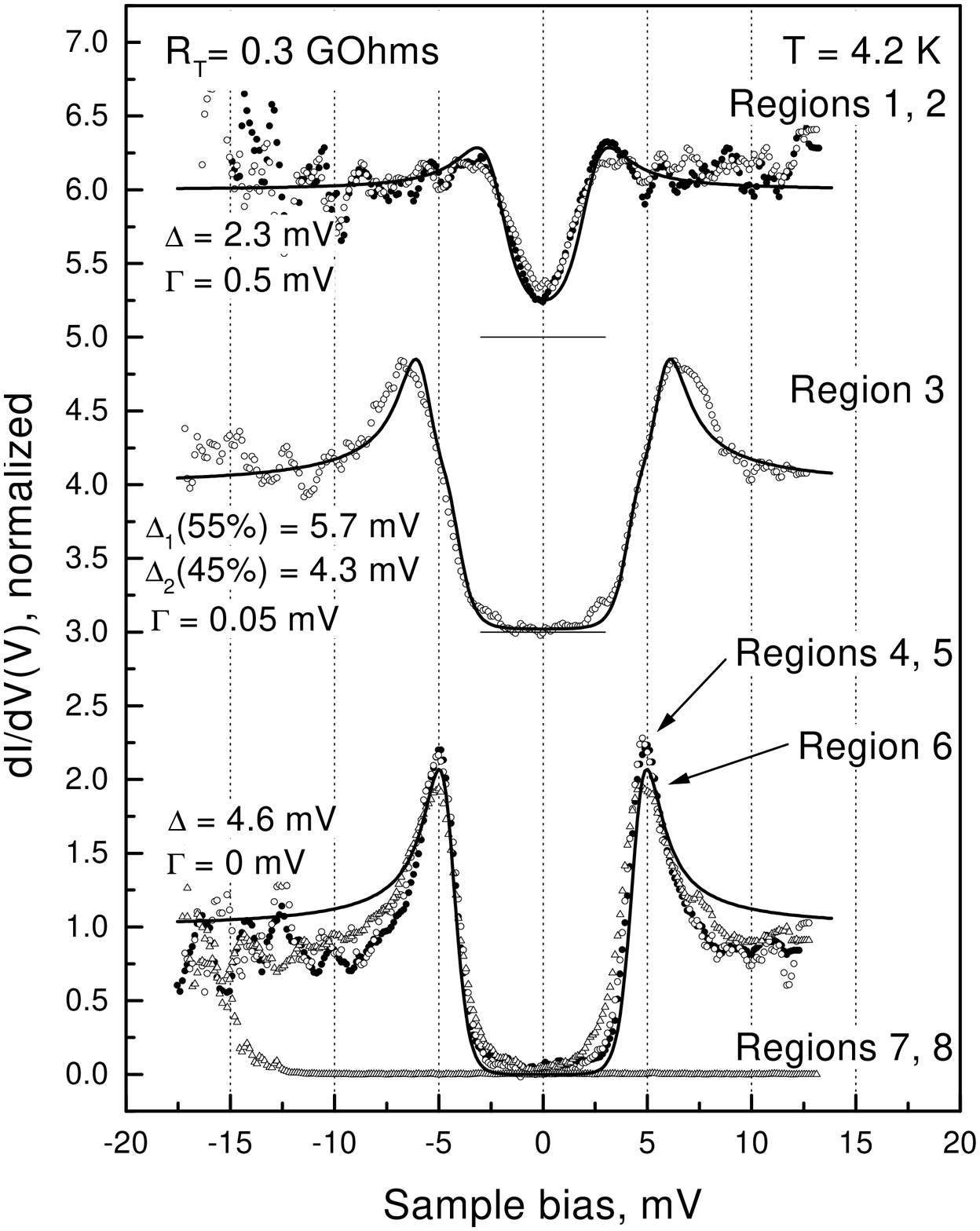}
\end{figure}

{\small FIG.4.  Normalized tunneling conductance spectra
corresponding to the regions 1-8. The data points are shifted for
clarity (horizontal bars indicate zero tunneling conductance). The
spectra in the regions 1-2 present a small gap $\Delta = 2.5\,$
meV. These smoother spectra are fit (solid line) with a large
$\Gamma$ ($0.3 \Delta$). In region 3, the gap width varies
spatially, giving a smeared averaged characteristic, well fit
(solid line) by two gaps of 4.3 mV and 5.7 mV and $\Gamma \ll
\Delta$.  In the regions 4-6 the spectra have a BCS shape with a
single gap.  The theoretical fit (solid line) with $\Delta =
4.6\,$meV and $\Gamma = 0\,$ meV is satisfactory. Regions 7-8:
tunneling conductance shows no states, i.e. insulating behavior,
in the energy window studied.}

\vskip 1mm

The detailed analysis of region 3 (about 20 nm size) shows that
two characteristic spectra are present: the first one, about 45\%
of the area, shows a gap of 4.3 meV, while the second one has a
gap of 5.7 meV (about 55\%).Thus, the average $dI/dV$ curve in
Fig.4, having a small $\Gamma = .05$ meV, has a kink at the small
gap energy. Regions 4-6 are the most homogeneous and have the
sharpest tunneling spectra with a clear single-gap BCS shape.  The
coherence peaks are very high and are situated at the same energy,
and no states exist inside. Only thermal broadening is needed
($\Gamma = 0$ meV), however the average gap value $\Delta = 4.6$
meV, is still significantly small, i.e. less than the ideal BCS
value of 5.9 meV by 20\%. In view of the double gap spectra, as in
Figs.1 and 2, this characteristic single gap must correspond to
the dirty limit. We conclude that about 99 \% of the STS
conductance spectra yield a single gap, with a probable reduced
$T_{C}$, due to proximity effects or the dirty limit cases. Thus,
the STS explains why the double gap DOS was not observed in
previous tunneling reports.

In conclusion, in this Letter we report the first direct evidence
for two superconducting gaps in MgB$_{2}$. These are due to
different electron-phonon coupling in the two distinct parts of
the Fermi surface, in agreement with the multiple gap model
\cite{Liu}. It is also consistent with the variation in relative
spectral weight observed, due to the arbitrary orientation of the
tunneling cone to the grains. The additional study of the
temperature dependence of the two gaps supports our conclusions
\cite{giubileo}. In the proximity effect or lowered surface
$T_{C}$ scenarios, the small gap should close at a lower
temperature, while it should survive until 39 K together with the
large gap in the double gap picture. No evidence for any large
anisotropy of the order parameter nor pair breaking is found in
the best spectra reported here. More significantly, the value of
the large gap ($7.5 \pm 0.5$ meV) leads to the ratio $2\Delta
/k_{B}T_{C} = 4.5 \pm 0.3$ indicating the presence of strong
electron-phonon coupling. STS performed on granular MgB$_{2}$ at
$T = 4.2 \,$K, shows the importance of spatial inhomogeneities in
such samples.

This work was supported by the Project ACI ``Nanostructures'' of
the French Ministry of Research.

\end{multicols}

\end{document}